\documentstyle[twoside, epsf]{article}

\input ibvs2.sty
\begin{document}
\IBVShead{5xxx}{00 Month 200x}

\IBVStitle{New R Coronae Borealis and DY Persei Candidates in the SMC}

\IBVSauth{Nikzat, F.$^{1,2}$; Catelan, M.$^{1,2}$}

\IBVSinst{Instituto de Astrof\'isica, Facultad de F\'isica, Pontificia Universidad Cat\'olica de Chile, Av. Vicu\~na Mackenna~4860, 782-0436 Macul, Santiago, Chile; e-mail: {\tt fnikzat@astro.puc.cl}, {\tt mcatelan@astro.puc.cl}}
\IBVSinst{Millennium Institute of Astrophysics, Santiago, Chile}

\IBVSkey{photometry}
\IBVSabs{We report 3 new R Coronae Borealis and 63 new DY~Persei candidates in the Small Magellanic Cloud. Our analysis, based on data published by the OGLE team, consisted in a search for the characteristic drops in brightness that define these classes. All candidates had been previously classified as semi-regular or Mira variables. We briefly remark upon the possible existence of a ``borderline'' DY~Per-like star and a ``transitional'' DY~Per/RCB stars. Follow-up observations are needed to conclusively establish the nature of our candidates.}

\begintext
R Coronae Borealis (RCB) stars are C-rich, H-deficient red supergiants that undergo dramatic dimming episodes at irregular intervals. The dimming episodes are caused by self-obscuration by dust, occurring as a consequence of mass loss events (e.g., Lambert \& Rao 1994; Clayton 1996, 2012; Catelan \& Smith 2015, and references therein). The purpose of this contribution is to report on a new search for RCB stars in the Small Magellanic Cloud (SMC), carried out using $VI$ light curves from the OGLE project (Soszy\'nski et al. 2011). To detect candidates, the $VI$ light curves of all SMC red variable stars were visually inspected, and compared against templates from the literature. New RCB candidates were detected in the process, which had previously been classified as semi-regular or Mira variables. Additionally, DY~Persei candidates were also identified. Compared to their RCB counterparts, the DY~Per stars tend to be cooler, have slower decline rates, and more symmetrical declines (e.g., Za{\v c}s et al. 2007; Catelan \& Smith 2015, and references therein). If confirmed, these detections would lead to a significant increase in the number of known RCB + DY~Per stars in the SMC.

The RCB stars have traditionally been classified as eruptive variables, due to their massive ejection episodes. However, they may also be classified as {\em self-eclipsing variable stars}, because of the self-obscuration due to formation of carbon dust around the star during mass loss events. Consequently, the RCB stars show a deep drop in their light curves which is a distinguishing characteristic of this class of variables. Since the dust forms in discrete clouds, the declines only occur when dust condenses along our line of sight. 

The evolutionary origin of the RCB stars is not understood yet. Two scenarios have been proposed for their formation (Iben et al. 1996; Saio \& Jeffery 2002): a merger of two white dwarfs or a final He-shell flash of the central object of a planetary nebula-hosting post-asymptotic giant branch (AGB) star. In the latter case, they would represent so-called ``born-again stars,'' to the extent that they would constitute (pre-) white dwarf stars that have been brought back to giant dimensions (Renzini 1990); in the former, they would be low-mass analogs of the same process that is believed to result in type Ia supernovae.

RCB stars are rare, with only about a hundred currently known (Tisserand et al. 2016), of which roughly one quarter are found in the Magellanic Clouds (Tisserand et al. 2013). To properly understand their evolutionary origin, more RCB stars in different environments with different metallicities are required. Furthermore, AGB stars are known as one of the main producers of dust to the interstellar medium (ISM), and likewise RCB stars may also significantly contribute to the dust enrichment of the ISM. As dust has different behavior in different environments, building significant samples of low-metallicity RCB stars can provide useful constraints on the role such stars may have played in the course of cosmic history. 

In this note, we present new RCB candidates found in the relatively low-metallicity environment of the SMC, based on an analysis of the morphology of the light curves of red variables published by the OGLE team. Their names and coordinates are provided in Table~1. To date, only three RCB and three DY~Per stars have been confirmed in the SMC (Tisserand et al. 2009), with an additional two RCB plus three DY~Per candidates also having been reported in the literature (Kraemer et al. 2005; Tisserand et al. 2009). Most recently, a new RCB candidate, Gaia16aau, was discovered using Gaia data (Tisserand et al. 2016). 

The catalog data for red variable stars in the SMC are available online from the OGLE project.\footnote{{\tt ftp://ftp.astrouw.edu.pl/ogle/ogle3/OIII-CVS/smc/lpv/}} The data were taken with the 1.3-meter Warsaw telescope at Las Campanas Observatory, northern Chile, in the course of the OGLE-III campaigns (Soszy\'nski et al. 2011). All $VI$ light curves were visually inspected, in a search for dramatic, non-periodic drops in brightness that might be indicative of RCB-like behavior. Our results are summarized in Table~1 and Figures~1 through 14. In total, we present two new RCB (Fig.~1) and 63 new DY~Per (Figs.~2-14) candidates. A third RCB candidate was also identified, and will be discussed later in this note. 

For completeness, previously confirmed and candidate RCB and DY~Per stars in the SMC are also listed in Table~2. Among these, three confirmed DY~Per stars (OGLE-SMC-LPV-03068, OGLE-SMC-LPV-04633, OGLE-SMC-LPV-11903) and three DY~Per candidates (OGLE-SMC-LPV-05023, OGLE-SMC-LPV-06616, OGLE-SMC-LPV-12291) from the EROS2 project (Tisserand et al. 2009) were detected in the OGLE data. Their corresponding light curves are shown in Figures~15 and 16 for the confirmed and candidate DY~Per stars, respectively. Note that we include OGLE-SMC-LPV-05007 among the DY~Per candidates in this paper, even though it was rejected by Tisserand et al. (2009) due to the presence of strong TiO bands. However, the light curve morphology bears some resemblance to those of other C-rich stars, and indeed the star has been classified as a C-star in the OGLE-III catalog. This star may thus be an interesting example of what might perhaps be called a ``borderline'' DY~Per-like star, not clearly conforming to the canonical DY~Per classification scheme. Its coordinates are given in the bottom row of Table~2. 

We emphasize, in this context, that OGLE-SMC-LPV-11903\footnote{Note that the OGLE-2 ID for this star is missing in the OGLE-III catalog.} (EROS2-SMC-DYPer-3), which has previously been classified as a DY~Per star based on spectroscopic data, presents a light curve morphology that is strongly reminiscent of an RCB star (Fig.~15). This may also hint at the possibility of a ``transitional'' DY~Per/RCB status. The latter might be consistent with the presence of an evolutionary sequence among hydrogen-deficient carbon stars, as suggested by Saio \& Jeffery (2002) and supported by De Marco et al. (2002) and Schaefer (2016). 

We also note that, while OGLE-SMC-LPV-03068 (EROS2-SMC-DYPer-2) is classified as an O-type LPV in the OGLE-III catalog, it has already been spectroscopically confirmed to be a DY~Per C-star (see Tisserand et al. 2004, 2009). The star's light curve, as shown in Figure~15, is indeed consistent with that expected for a C-star.  

We were able to match the spectroscopic RCB candidate MSX-SMC-014 (Kraemer et al. 2005; Tisserand et al. 2009) to OGLE-SMC-LPV-05719; the light curve is shown in Figure~17. We point out that this light curve bears some resemblance to that of OGLE-SMC-LPV-17611; in both cases, we see several photometric declines during the OGLE-III monitoring, and the time interval between adjacent minima/maxima is roughly similar (Fig.~17). We accordingly propose OGLE-SMC-LPV-17611 as an additional candidate RCB star in the SMC.

To close, we note that, in our work, we have only used light curve morphology as indicative of potential RCB/DY~Per status. Follow-up observations, both photometric and spectroscopic, are required in order to conclusively establish the nature of our candidates. 

\vskip 1cm
\noindent {\em Acknowledgments}: We warmly thank the referee, E. J. Montiel, for his thoughtful report. Support for this project is provided by the Ministry for the Economy, Development, and Tourism's Millennium Science Initiative through grant IC\,120009, awarded to the Millennium Institute of Astrophysics (MAS); by Proyecto Basal PFB-06/2007; by CONICYT's PCI program through grant DPI20140066; and by FONDECYT grants \#1141141. FN is grateful for financial support by Proyecto Gemini CONICYT grants \#32130013 and \#32140036.

\references
Catelan, M., Smith, H. A., 2015, {\em Pulsating Stars} (Wiley-VCH)

Clayton, G.~C.\ 1996, {\it PASP}, {\bf 108}, 225 

Clayton, G.~C., 2012, {\it JAVSO}, {\bf 40}, 539  

De Marco, O., Clayton, G.~C., Herwig, F., et al.\ 2002, {\it AJ}, {\bf 123}, 3387

Kraemer, K.~E., Sloan, G.~C., Wood, P.~R., et al., 2005, {\it ApJ}, {\bf 631}, {L147}   

Lambert, D.~L., \& Rao, N.~K.\ 1994, {\it JApA}, {\bf 15}, 47 

Renzini, A., 1990, {\it ASPC}, {\bf 11}, 549

Saio, H., \& Jeffery, C.~S.\ 2002, {\it MNRAS}, {\bf 333}, 121

Schaefer, B.~E.\ 2016, {\it MNRAS}, {\bf 460}, 1233 

Soszy{\'n}ski, I., Udalski, A., Szyma{\'n}ski, M.~K., et al., 2011, {\it AcA}, {\bf 61}, 217

Tisserand, P., Clayton, G.~C., Welch, D.~L., et al., 2013, {\it A\&A}, {\bf 551}, A77   

Tisserand, P., Marquette, J.~B., Beaulieu, J.~P., et al., 2004, {\it A\&A}, {\bf 424}, 245

Tisserand, P., Wood, P.~R., Marquette, J.~B., et al., 2009, {\it A\&A}, {\bf 501}, 985

Tisserand, P., Wyrzykowski, L., Clayton, G., et al., 2016, {\it ATel}, {\bf 8681}, 1T  

Za{\v c}s, L., Mondal, S., Chen, W.~P., et al.\ 2007, {\it A\&A}, {\bf 472}, 247 

\endreferences


\newpage
\vskip 1cm
\centerline{Table 1. New RCB and DY~Per Candidates in the SMC}
\begin{center}
\begin{tabular}{ccccc}
\hline
\hline
Star name   & 	 RA (J2000)	& DEC (J2000) & Other ID & Subtype  \\
\hline
\multicolumn{5}{c}{New RCB Candidates ({\em this note})}\\
\hline
OGLE-SMC-LPV-01019 &    00:31:16.77  & -73:56:48.6   & 			    & SRV 	 \\
OGLE-SMC-LPV-06216 &    00:47:05.33  & -72:34:30.5   & MACHO-208.15801.1    & SRV   	\\
OGLE-SMC-LPV-17611 &    01:09:21.69  & -71:24:35.1   &  		    & Mira \\ 	
\hline
\multicolumn{5}{c}{New DY~Per Candidates ({\em this note})}\\
\hline
OGLE-SMC-LPV-00190 &    00:23:58.50  & -73:37:54.9   &			    & Mira	\\ 
OGLE-SMC-LPV-00486&	00:27:10.92  & -73:24:30.3   & 			    & Mira	\\
OGLE-SMC-LPV-00492&     00:27:14.50  & -73:36:42.4   &			    & Mira	\\ 
OGLE-SMC-LPV-00666&     00:28:34.78  & -74:05:41.9   &                 	    & SRV 	\\       
OGLE-SMC-LPV-00799 & 	00:29:42.22  & -73:19:11.1   & 			    & SRV	\\

OGLE-SMC-LPV-02715 &    00:39:13.28  & -73:57:05.9   &			    & SRV	\\ 

OGLE-SMC-LPV-03078 &    00:40:16.26  & -73:01:15.5   & 			    & Mira	\\ 
OGLE-SMC-LPV-03315 &    00:40:55.61  & -72:47:49.4   & 			    & Mira      \\ 
OGLE-SMC-LPV-03429&	00:41:16.95  & -72:52:16.8   & MACHO-213.15398.77   & Mira      \\ 
OGLE-SMC-LPV-03593 & 	00:41:42.45  & -72:58:53.7   & MACHO-213.15453.892  & Mira      \\	
OGLE-SMC-LPV-03810 & 	00:42:16.10  & -72:57:32.6   & 			    & SRV       \\ 

OGLE-SMC-LPV-04208&	00:43:09.58  & -73:09:20.1   &  		    & Mira      \\ 
OGLE-SMC-LPV-04575&    00:43:56.94  &  -73:56:08.0   &			    & Mira      \\  
   	
OGLE-SMC-LPV-05322 &   00:45:33.00  & -73:05:12.1    & MACHO-212.15679.930  & Mira       \\   
OGLE-SMC-LPV-05801  &   00:46:23.52  & -72:37:39.8   & 			    & Mira       \\ 

OGLE-SMC-LPV-06089 & 	00:46:50.78  & -71:47:39.4   &			    & SRV        \\
OGLE-SMC-LPV-06156 & 	00:46:59.11  & -73:25:18.8   & MACHO-212.15788.31   & SRV   	\\ 
OGLE-SMC-LPV-06572 & 	00:47:37.82  & -73:00:13.3   & MACHO-212.15795.25   & SRV   	\\ 
OGLE-SMC-LPV-06962 & 	00:48:12.30  & -72:41:18.6   &  		    & SRV   	\\ 

OGLE-SMC-LPV-07113&    00:48:27.01  & -72:45:55.5   & MACHO-208.15855.5029 & Mira  	\\ 
OGLE-SMC-LPV-07354 & 	00:48:50.92  & -73:14:02.3   &  		    & SRV   	\\   
OGLE-SMC-LPV-07375&	00:48:52.49  & -73:08:56.8   & MACHO-212.15907.28   & Mira  	\\    
OGLE-SMC-LPV-07665&    00:49:20.53  & -72:34:11.8   & MACHO-208.15915.2828 & Mira  	\\  
OGLE-SMC-LPV-07829 & 	00:49:34.97  & -73:18:18.5   & MACHO-212.15904.2217 & SRV   	\\ 

OGLE-SMC-LPV-08192 & 	00:50:12.59  & -72:33:42.3   & MACHO-208.15972.2547 & SRV  	\\   
OGLE-SMC-LPV-08390 &    00:50:31.29  & -72:29:13.1   & MACHO-208.15973.50   & Mira		\\  
OGLE-SMC-LPV-08445 &    00:50:37.00  & -73:08:53.7   & 			    & SRV	\\ 
OGLE-SMC-LPV-08741 & 	00:51:04.65  & -72:01:37.6   & 		 	    & SRV 	\\
OGLE-SMC-LPV-08803&	00:51:10.37  & -72:27:42.8   &			    & Mira	\\ 
OGLE-SMC-LPV-08931 &    00:51:23.06  & -72:36:16.4   & 			    & SRV	\\ 

OGLE-SMC-LPV-09350 &    00:51:58.14  & -73:43:35.3   &			    & SRV	\\ 
OGLE-SMC-LPV-09801&	00:52:40.18  & -72:47:27.7   & MACHO-207.16140.490  & Mira	\\      

OGLE-SMC-LPV-10280 & 	00:53:23.12  & -72:04:22.4   &  		    & SRV	\\
OGLE-SMC-LPV-10436 &    00:53:37.31  & -72:34:35.2   &  MACHO-207.16200.324  & Mira	\\  
OGLE-SMC-LPV-10465 & 	00:53:40.01  & -72:52:18.7   &  		    & SRV	\\  
OGLE-SMC-LPV-10816 &    00:54:10.75  & -73:03:03.1   & 	MACHO-211.16250.24  & SRV	\\ 
OGLE-SMC-LPV-11279&	00:54:54.11  & -73:03:18.2   & MACHO-211.16250.4090 & Mira  	\\ 
OGLE-SMC-LPV-11698 &    00:55:34.14  & -72:40:29.4   &  MACHO-207.16313.24  & SRV	\\   
OGLE-SMC-LPV-11806&    00:55:44.47  & -72:54:40.8   &   		    & Mira	\\ 

OGLE-SMC-LPV-12043 &    00:56:10.05  & -72:28:41.9   & 			    & Mira \\	 
OGLE-SMC-LPV-12119 &    00:56:16.38  & -72:16:41.4   & MACHO-207.16376.687  & Mira \\	 

\hline
\hline
\end{tabular}
\end{center}
\vskip 1cm


\newpage
\vskip 1cm
\centerline{Table 1. New RCB and DY~Per Stars in the SMC ({\em cont.})}
\vskip 3mm
\begin{center}
\begin{tabular}{ccccc}
\hline
\hline
Star name   & 	 RA (J2000)	& DEC (J2000)  &  Other ID & Subtype \\  
\hline
\multicolumn{5}{c}{New DY~Per Candidates ({\em this note})}\\
\hline

OGLE-SMC-LPV-12304 & 	00:56:36.77  & -73:32:55.5   & 			    & SRV  \\	
OGLE-SMC-LPV-12427 &    00:56:50.29  & -72:25:08.7   & MACHO-207.16373.675  & Mira \\	
OGLE-SMC-LPV-13205 & 	00:58:20.78  & -72:55:02.1   & MACHO-211.16480.1110 & SRV \\	
OGLE-SMC-LPV-13251 & 	00:58:26.59  & -73:40:35.0   &  		    & SRV \\	
OGLE-SMC-LPV-13320 & 	00:58:34.86  & -73:32:10.9   & 			    & SRV \\	

OGLE-SMC-LPV-13323 &    00:58:35.18  & -72:59:35.6   & MACHO-211.16479.2    & SRV \\ 
OGLE-SMC-LPV-13676 &    00:59:15.78  & -72:27:54.6   & MACHO-207.16544.36   & Mira \\	                      
OGLE-SMC-LPV-13739 &  	00:59:21.90  & -72:11:13.4   &   		    & SRV \\ 
OGLE-SMC-LPV-13749 &    00:59:23.36  & -73:56:01.0   &			    & SRV  \\	

OGLE-SMC-LPV-14197 &    01:00:15.67  & -72:22:26.2   & MACHO-207.16602.106  & SRV \\ 
OGLE-SMC-LPV-14205 &    01:00:16.84  & -72:55:18.1   &  		    & Mira \\	
OGLE-SMC-LPV-14322 &    01:00:31.66  & -72:14:49.1   & MACHO-207.16604.926  & Mira \\	 
OGLE-SMC-LPV-14778 &    01:01:26.59  & -72:47:41.2   & 			    & Mira \\	         
OGLE-SMC-LPV-14991 &    01:01:54.59  & -72:58:22.4   & MACHO-211.16707.28   & Mira \\ 	

OGLE-SMC-LPV-16113 &    01:04:39.84  & -72:49:48.2   &  		    & Mira \\	
OGLE-SMC-LPV-16844 &	01:06:53.07  & -73:46:00.2   &  		    & Mira \\	
OGLE-SMC-LPV-16850 &    01:06:54.81  & -72:24:41.2   &			    & Mira \\	

OGLE-SMC-LPV-17194 & 	01:08:01.14  & -72:53:17.4   &  		    & SRV \\	
OGLE-SMC-LPV-17267 &    01:08:12.97  & -72:52:44.0   & 		       	    & Mira \\	 
OGLE-SMC-LPV-17976 & 	01:10:53.22  & -72:14:46.0   &  		    & SRV \\	

OGLE-SMC-LPV-18657 &    01:15:09.88  & -72:05:51.3   &  		    & Mira \\	
OGLE-SMC-LPV-19032 &    01:18:48.06  & -72:27:43.7   &			    & Mira \\	  
\hline
\hline
\end{tabular}
\end{center}
\vskip 1cm


\vfill\eject
\centerline{Table 2. Known RCB and DY~Per Stars in the SMC}
\vskip 3mm
\begin{center}
\begin{tabular}{cccc}
\hline
\hline
Star name   & 	 RA (J2000)	& DEC (J2000)  &  Other ID\\
\hline
\multicolumn{4}{c}{ RCB and DY~Per Confirmed }\\
\hline
EROS2-SMC-RCB-1 &		00:37:47.11  & -73:39:02.3   & RAW-21 \\
EROS2-SMC-RCB-2 &		00:48:22.96  & -73:41:04.7   & RAW-476 \\
EROS2-SMC-RCB-3 & 		00:57:18.15  & -72:42:35.2   & MACHO-207.16426.1662 \\
EROS2-SMC-DYPer-1 & 		00:44:07.50  & -72:44:16.4   &  RAW-233 \\ 
				  & 					  & 			  & MACHO-208.15571.60 \\
                  &	                     &               &  OGLE-SMC-LPV-04633 \\

EROS2-SMC-DYPer-2 & 		00:40:14.72  & -74:11:21.6   &  [MH95]-431 \\
                  &             	     &               &  OGLE-SMC-LPV-03068 \\

EROS2-SMC-DYPer-3 &             00:55:54.97  & -72:35:12.27  &  RAW-961    \\
&  			     &  	     &  		  OGLE2-SMC-SC7-368043\\
		  &  			     &  	     &  OGLE-SMC-LPV-11903 \\ 
\hline
\multicolumn{4}{c}{ RCB and DY~Per Candidates }\\
\hline
EROS2-SMC-RCB-4   &    01:04:52.89  & -72:04:02.64   &   OGLE2-SMC-SC10-107856 \\ 
MSX-SMC-014       &    00:46:16.33  & -74:11:13.6    &   OGLE-SMC-LPV-05719  \\
Gaia16aau         &    00:50:10.67   & -69:43:57.9   &   [MH95]-580 \\
                  &                 &                &   OGLE-SMC710.08.1\\

EROS2-SMC-DYPer-4 &    00:56:35.47  & -71:32:32.66   &   [MH95]-672			\\	
		  &   		    &		     &   OGLE-SMC-LPV-12291 \\

EROS2-SMC-DYPer-5 &    00:47:41.71  & -73:06:16.38   &   RAW-421 \\ 
				&					& 				& MACHO-212.15793.25 \\ 
		  &   		    &		     &   OGLE-SMC-LPV-06616  \\

EROS2-SMC-DYPer-6 &    00:44:56.40   & -73:12:25.02  &   MACHO-212.15621.153 \\	
		  &		     &		     &	OGLE-SMC-LPV-05023    \\
\hline
\multicolumn{4}{c}{``Borderline'' DY~Per-Like Candidate ({\em see text})}\\
\hline
sm0101n-16084   &	00:44:54.02   & -73:15:30.02   & MACHO-212.15620.713 \\
		&	  	      & 	       & OGLE-SMC-LPV-05007 \\
\hline
\hline
\end{tabular}
\end{center}
\vskip 1cm

\IBVSfig{8cm}{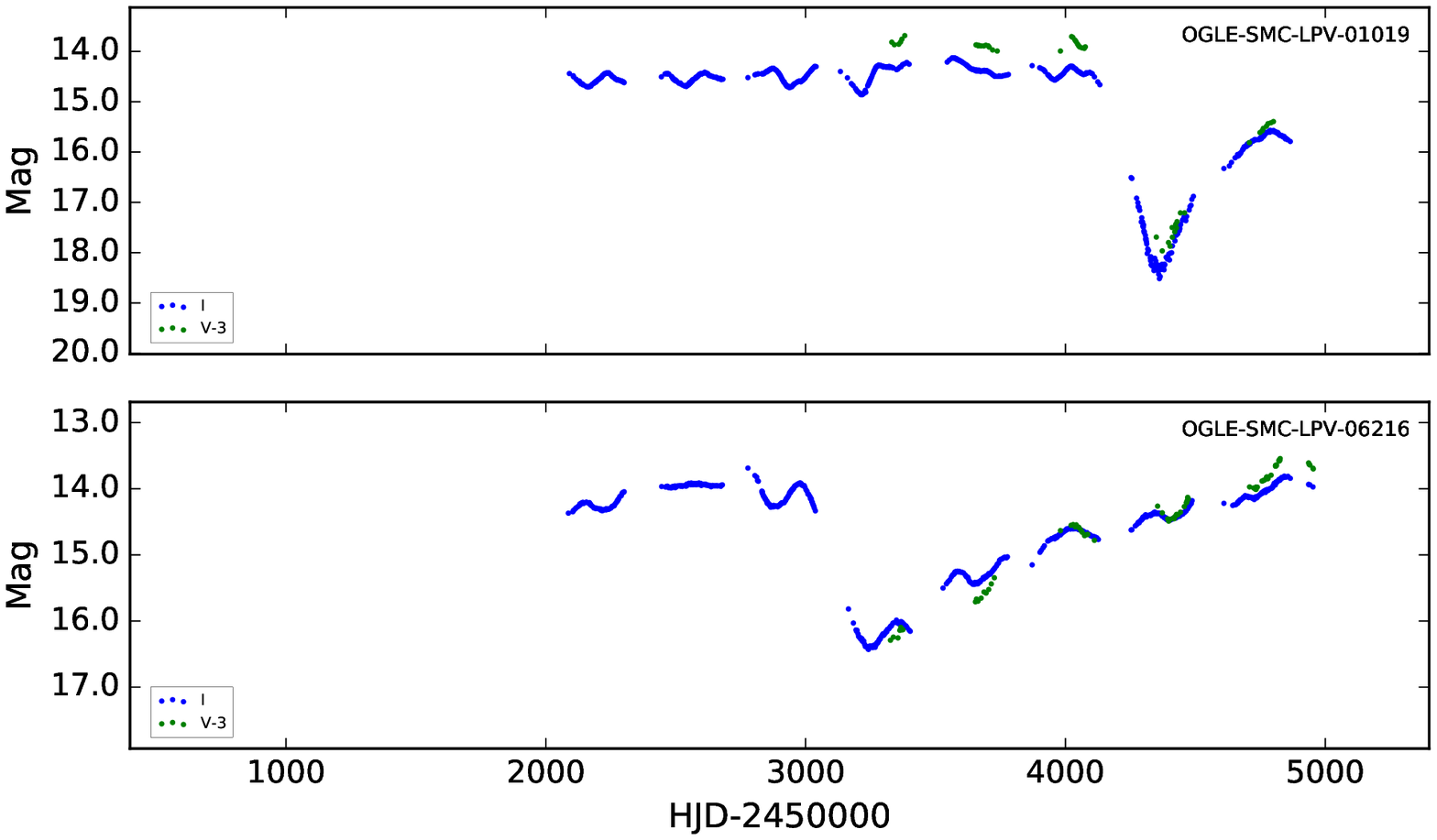}{Light curves in $I$ ({\em blue}) and $V$ ({\em green}) of two new RCB candidates in the SMC.}
\label{fig:fig1}
\vskip 1cm
\IBVSfig{20cm}{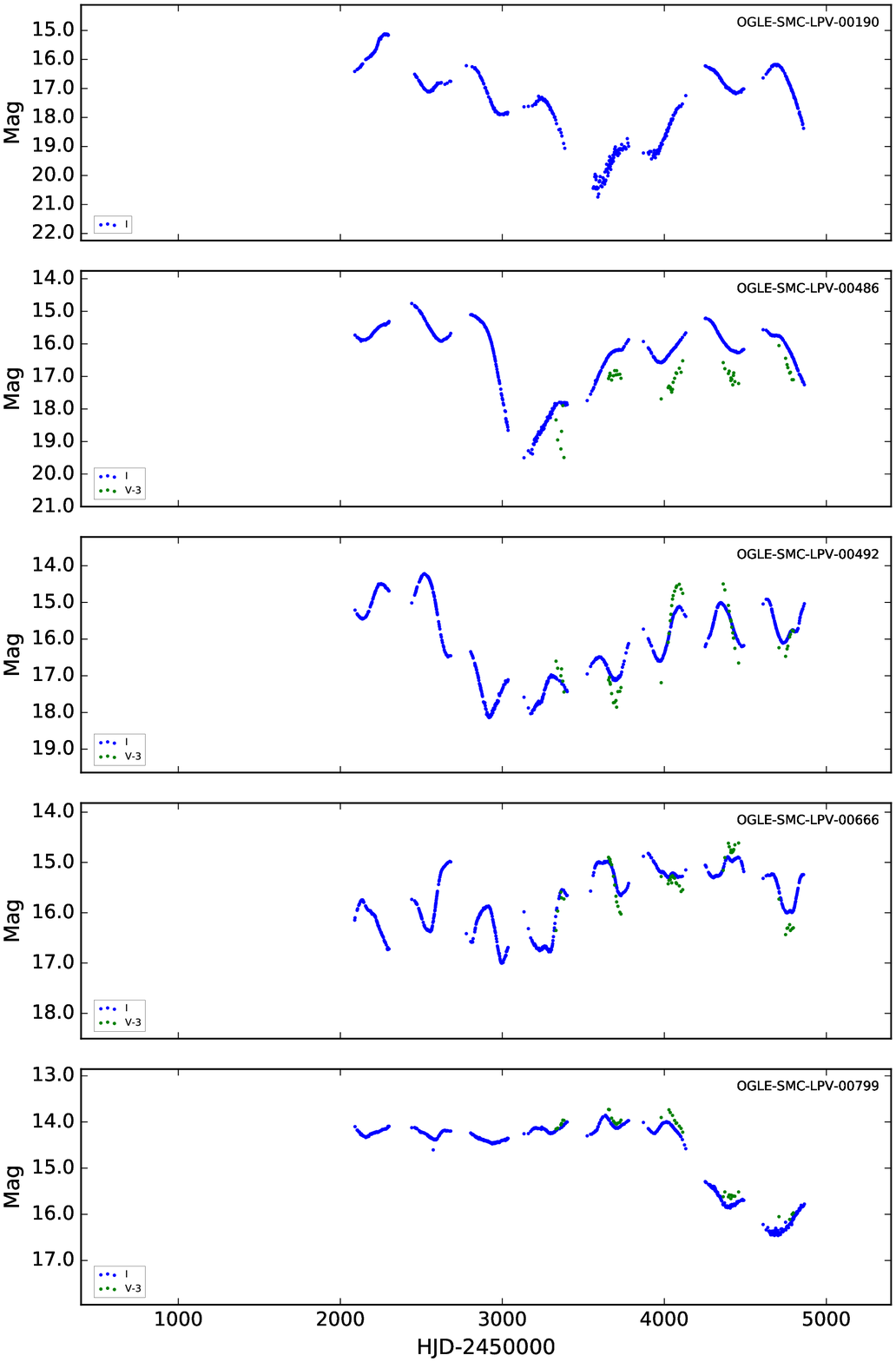}{Light curves in $I$ ({\em blue}) and $V$ ({\em green}) of new DY~Per candidates in the SMC.}
\label{fig:fig2}
\vskip 1cm

\IBVSfig{20cm}{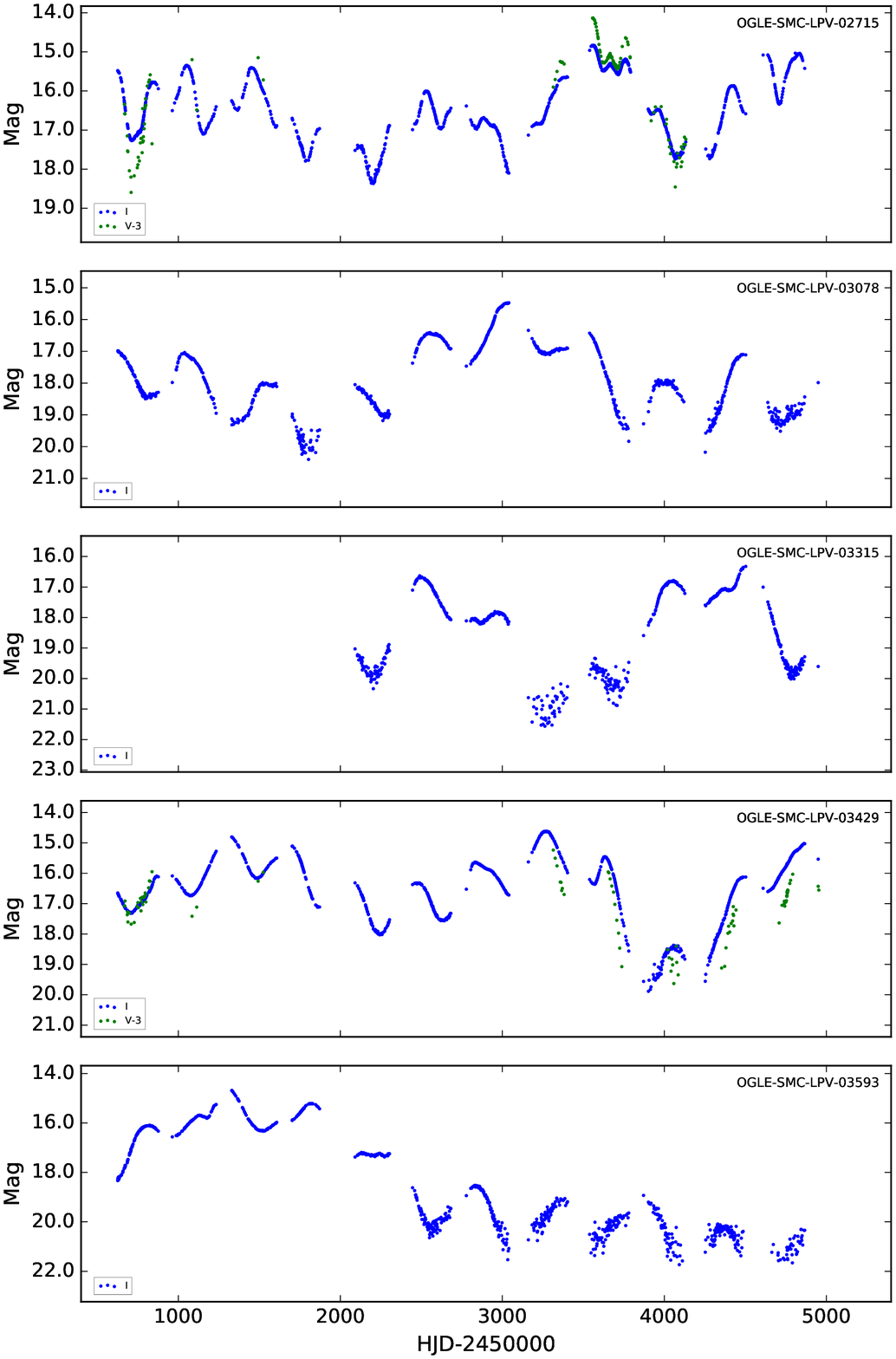}{Light curves in $I$ ({\em blue}) and $V$ ({\em green}) of new DY~Per candidates in the SMC.}
\label{fig:fig3}
\vskip 1cm

\IBVSfig{20cm}{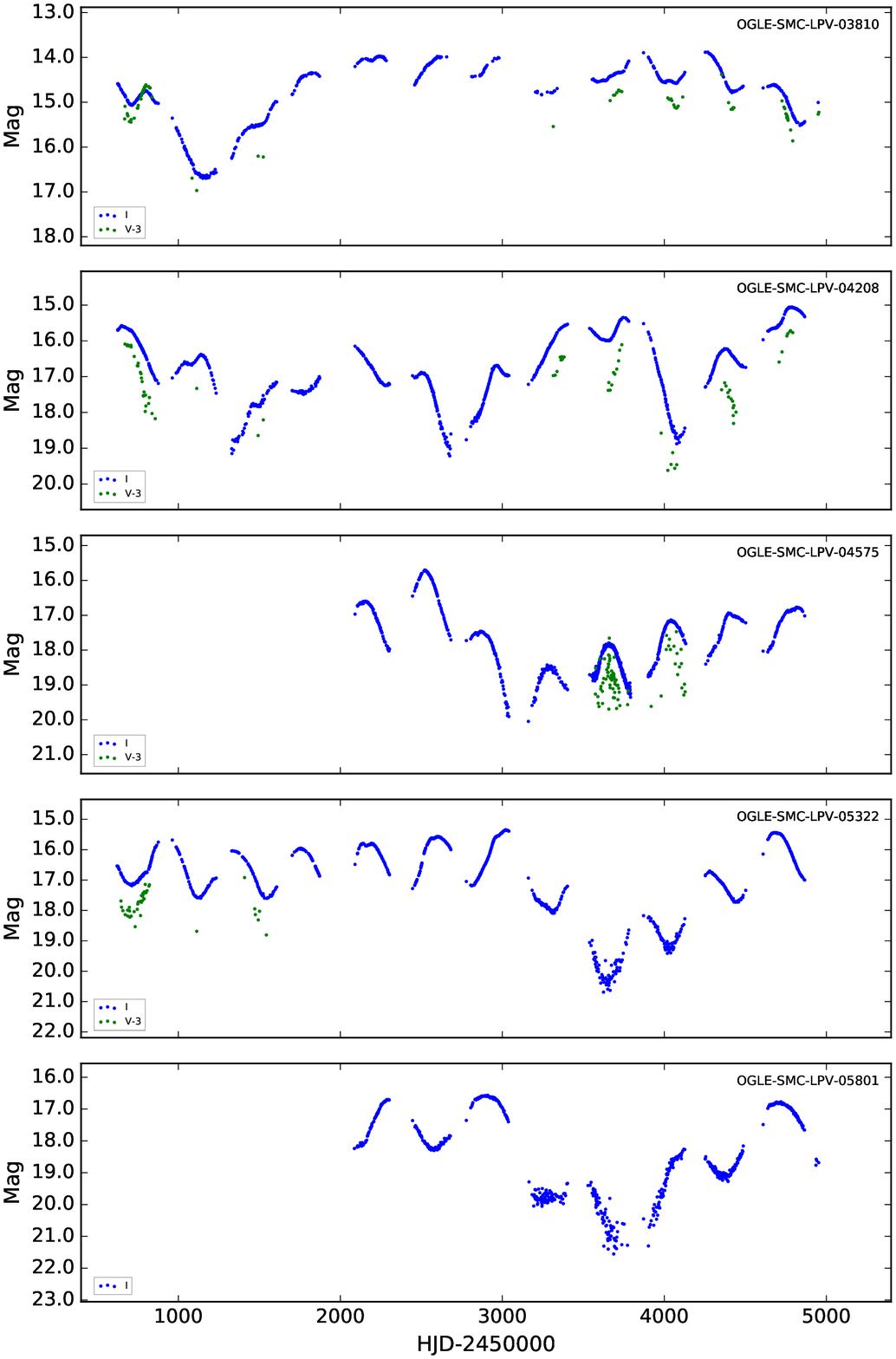}{Light curves in $I$ ({\em blue}) and $V$ ({\em green}) of new DY~Per candidates in the SMC.}
\label{fig:fig4}
\vskip 1cm

\IBVSfig{20cm}{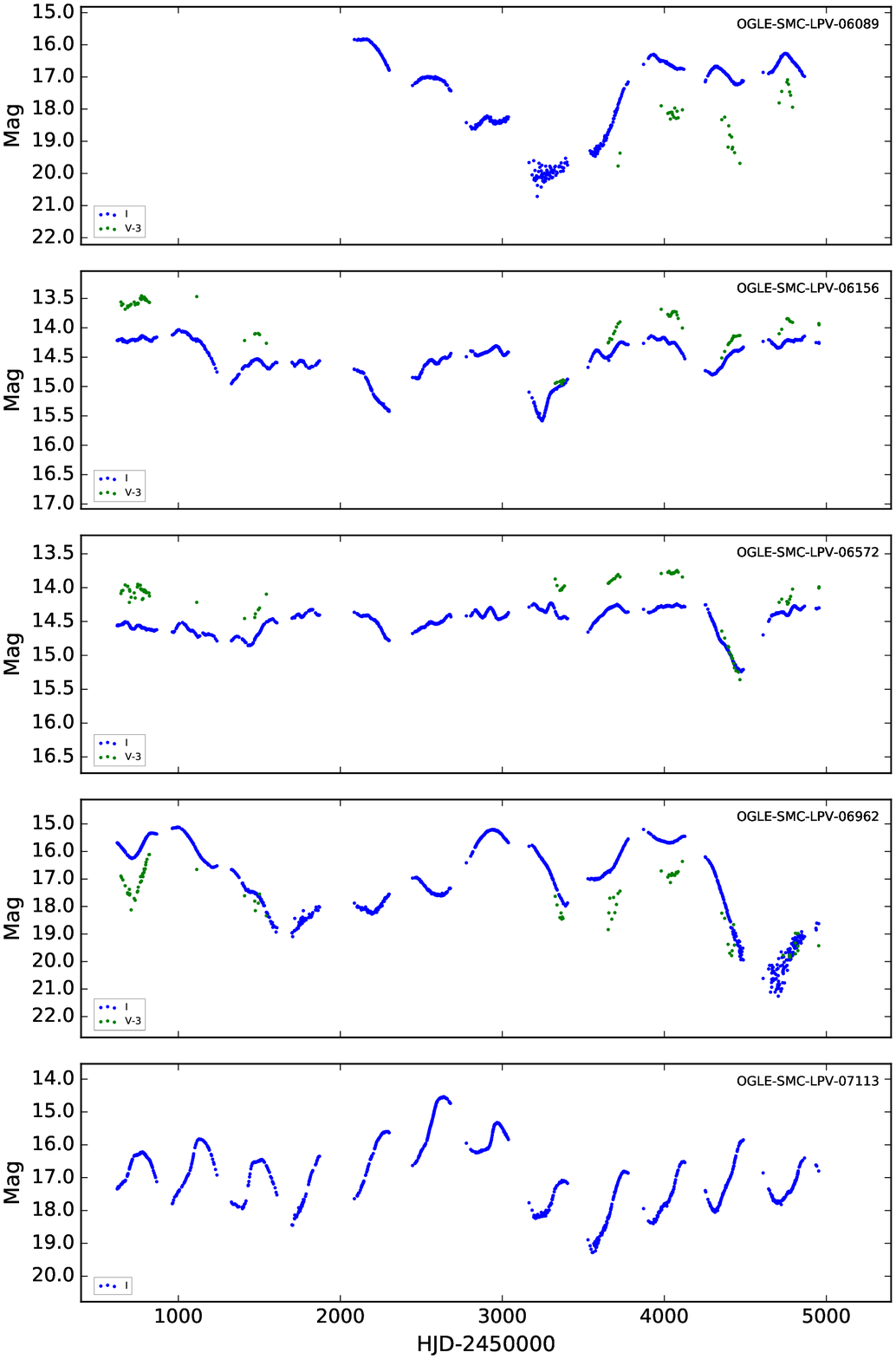}{Light curves in $I$ ({\em blue}) and $V$ ({\em green}) of new DY~Per candidates in the SMC.}
\label{fig:fig5}
\vskip 1cm

\IBVSfig{20cm}{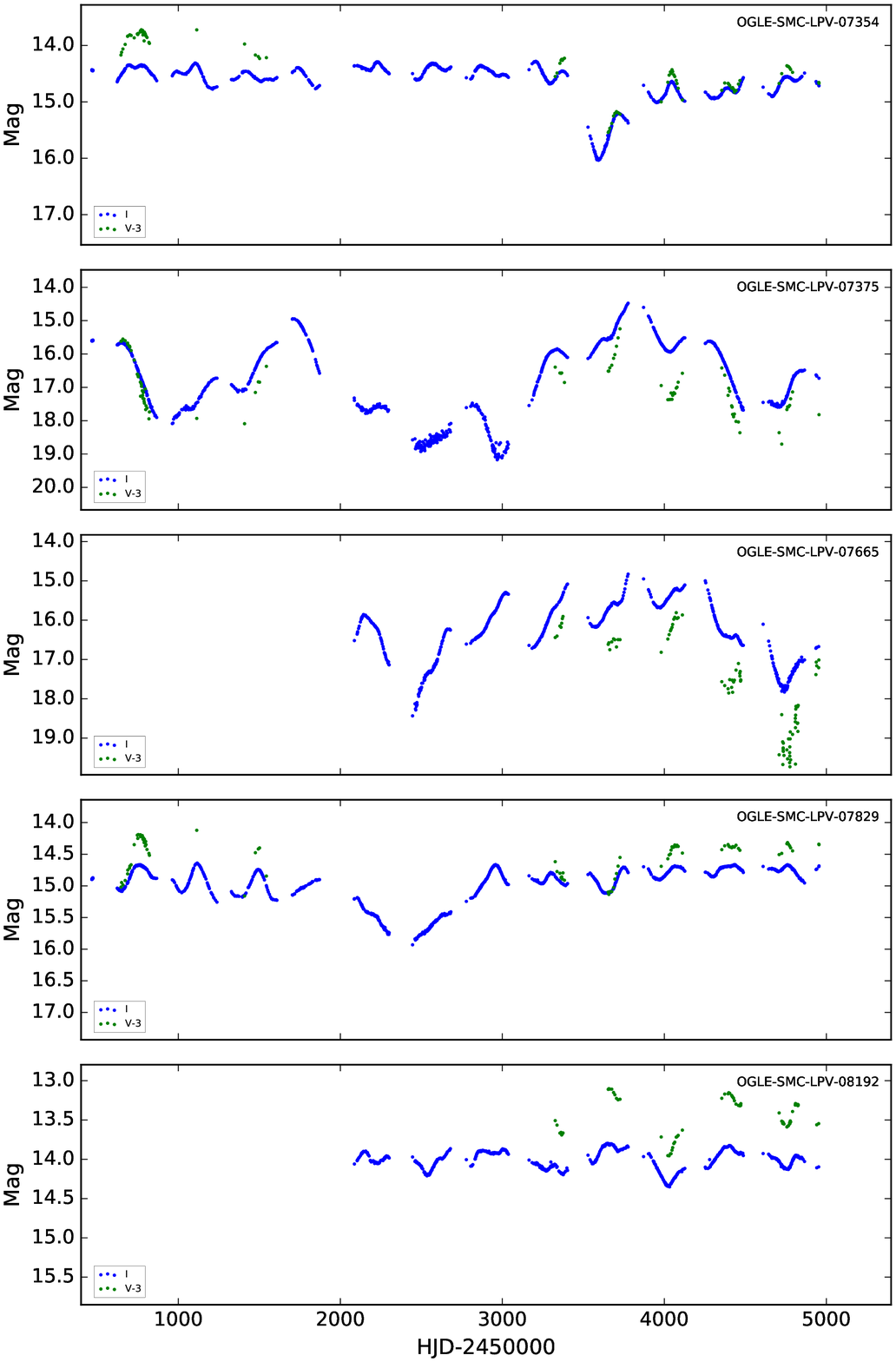}{Light curves in $I$ ({\em blue}) and $V$ ({\em green}) of new DY~Per candidates in the SMC.}
\label{fig:fig6}
\vskip 1cm

\IBVSfig{20cm}{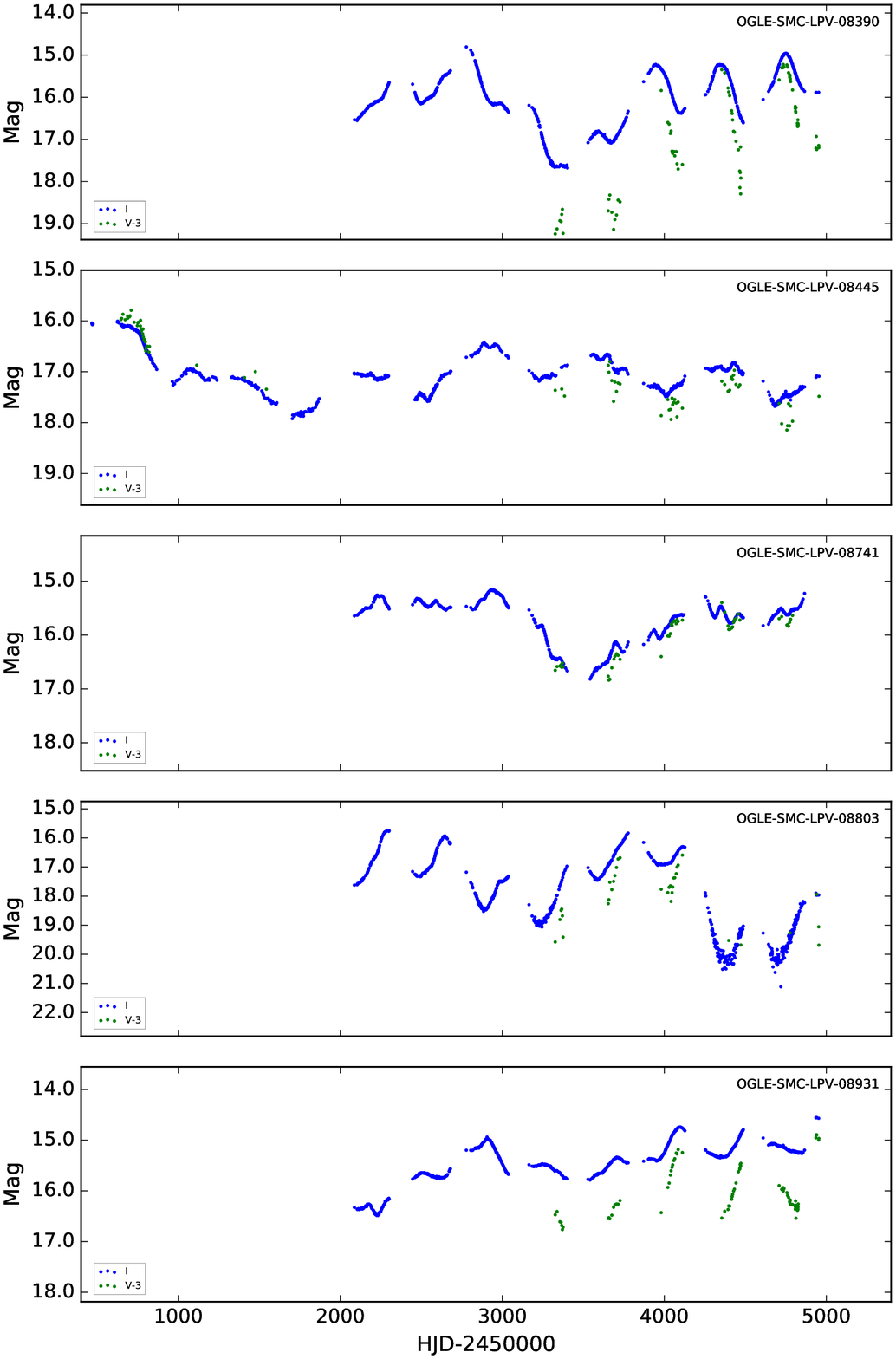}{Light curves in $I$ ({\em blue}) and $V$ ({\em green}) of new DY~Per candidates in the SMC.}
\label{fig:fig7}
\vskip 1cm

\IBVSfig{20cm}{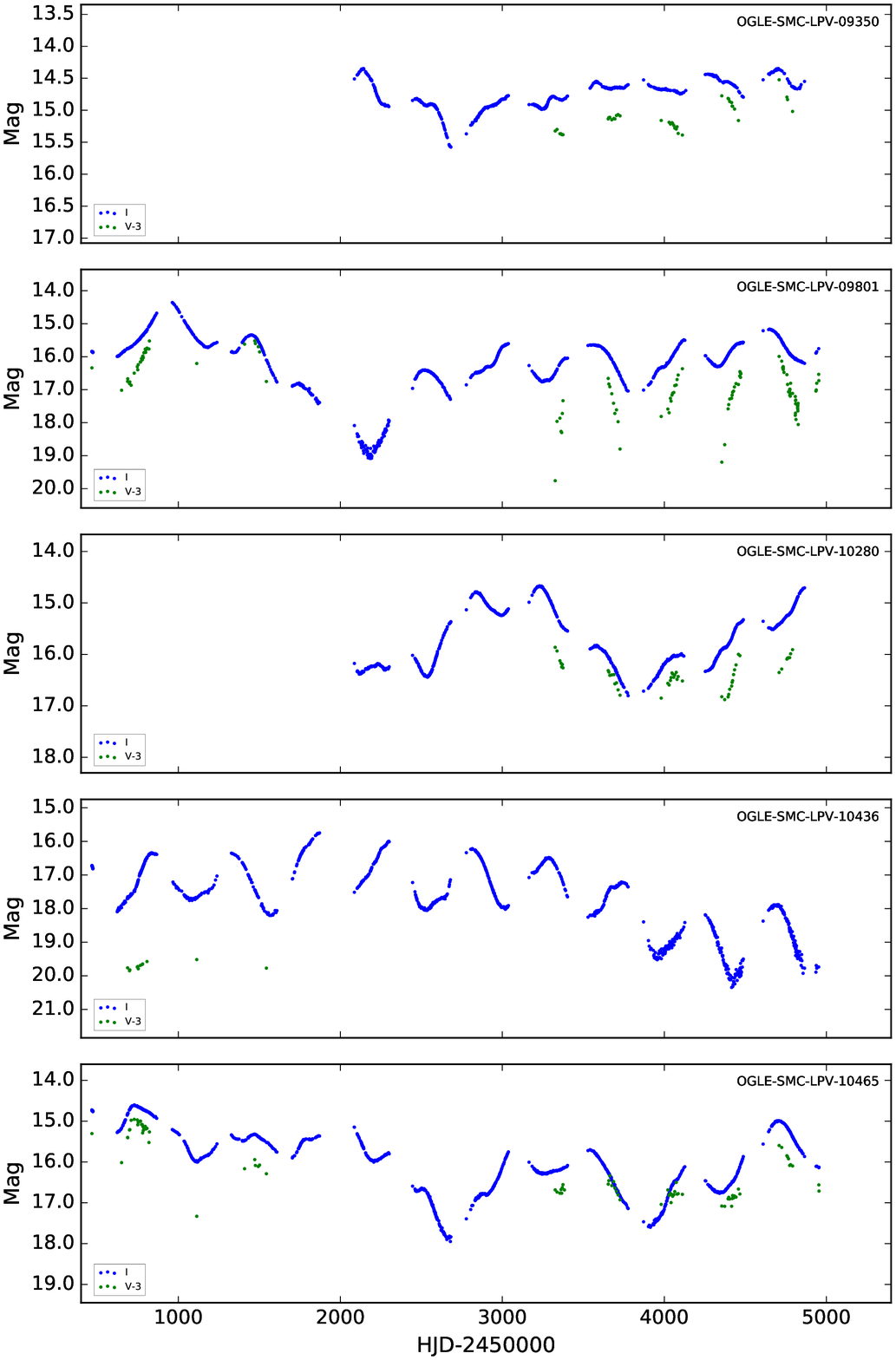}{Light curves in $I$ ({\em blue}) and $V$ ({\em green}) of new DY~Per candidates in the SMC.}
\label{fig:fig8}
\vskip 1cm

\IBVSfig{20cm}{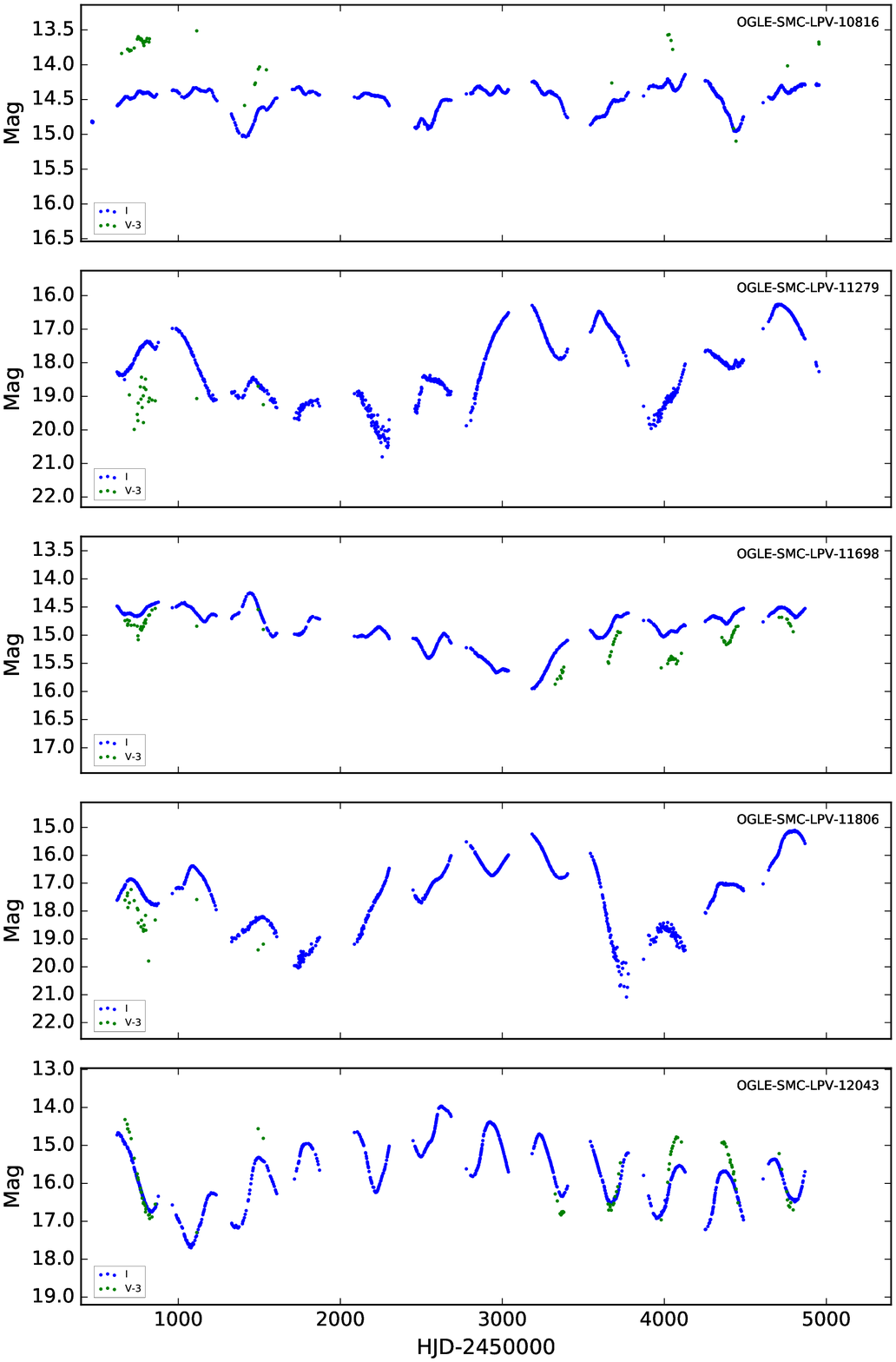}{Light curves in $I$ ({\em blue}) and $V$ ({\em green}) of new DY~Per candidates in the SMC.}
\label{fig:fig9}
\vskip 1cm

\IBVSfig{20cm}{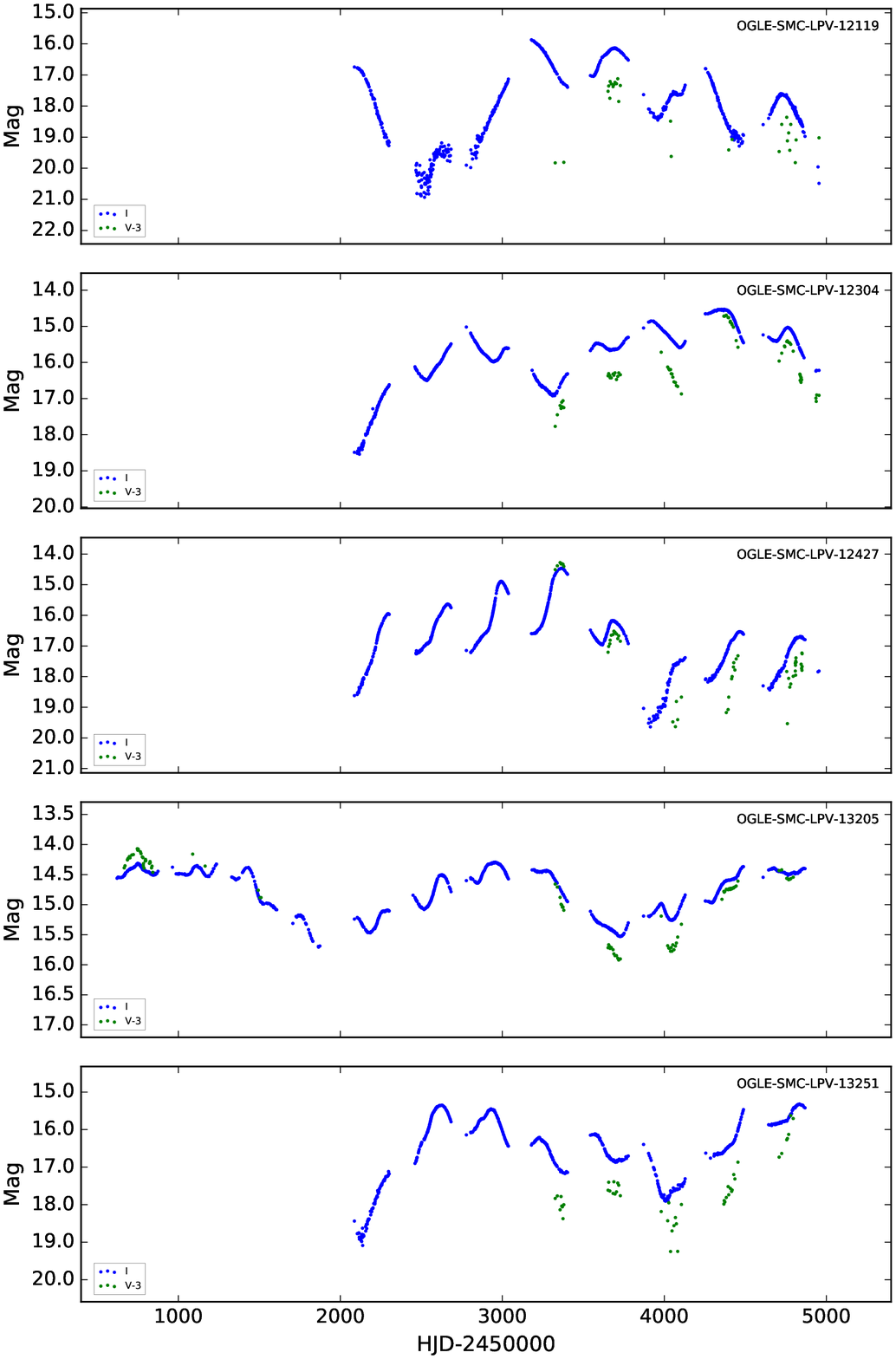}{Light curves in $I$ ({\em blue}) and $V$ ({\em green}) of new DY~Per candidates in the SMC.}
\label{fig:fig10}
\vskip 1cm

\IBVSfig{20cm}{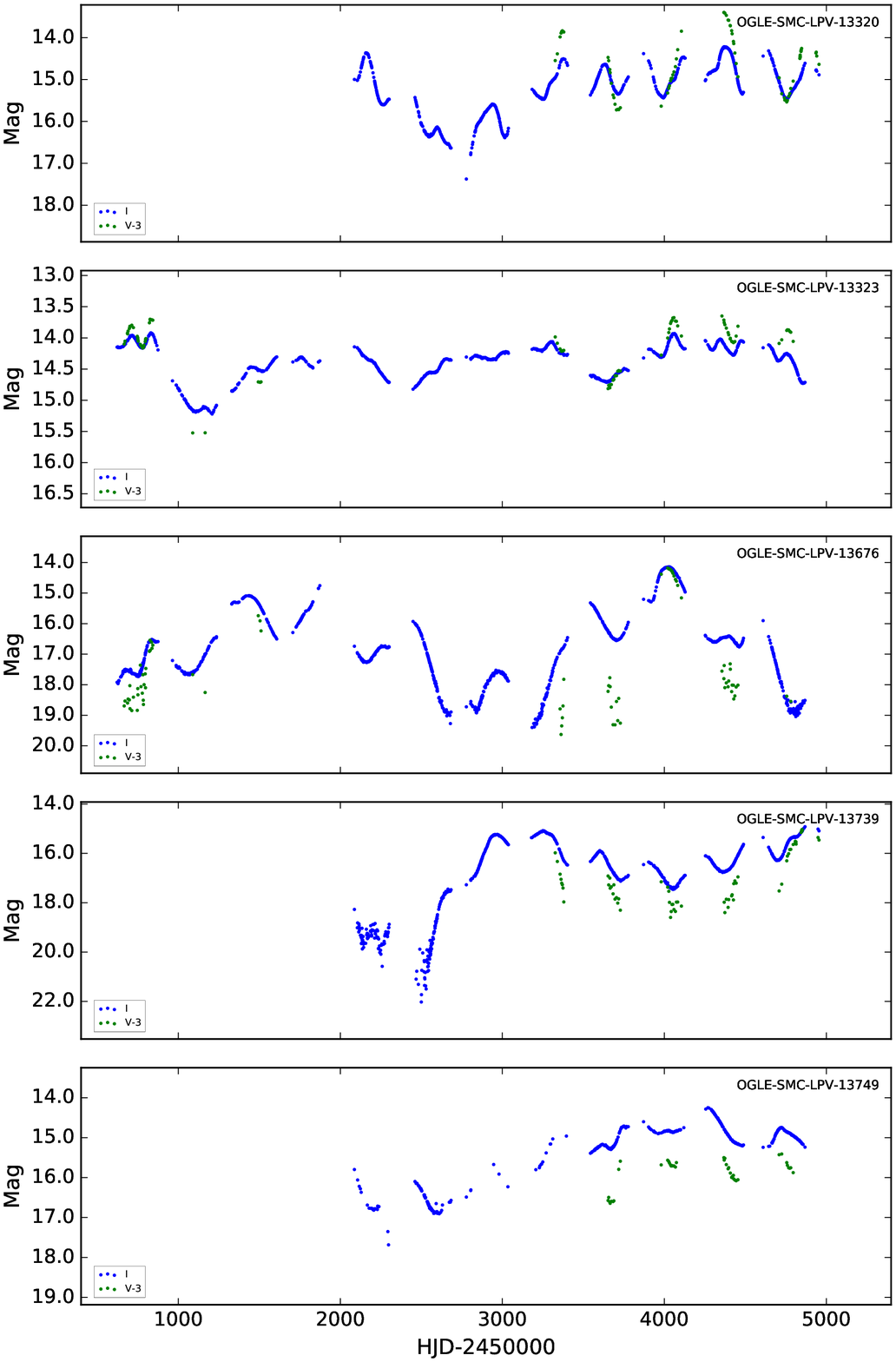}{Light curves in $I$ ({\em blue}) and $V$ ({\em green}) of new DY~Per candidates in the SMC.}
\label{fig:fig11}
\vskip 1cm

\IBVSfig{20cm}{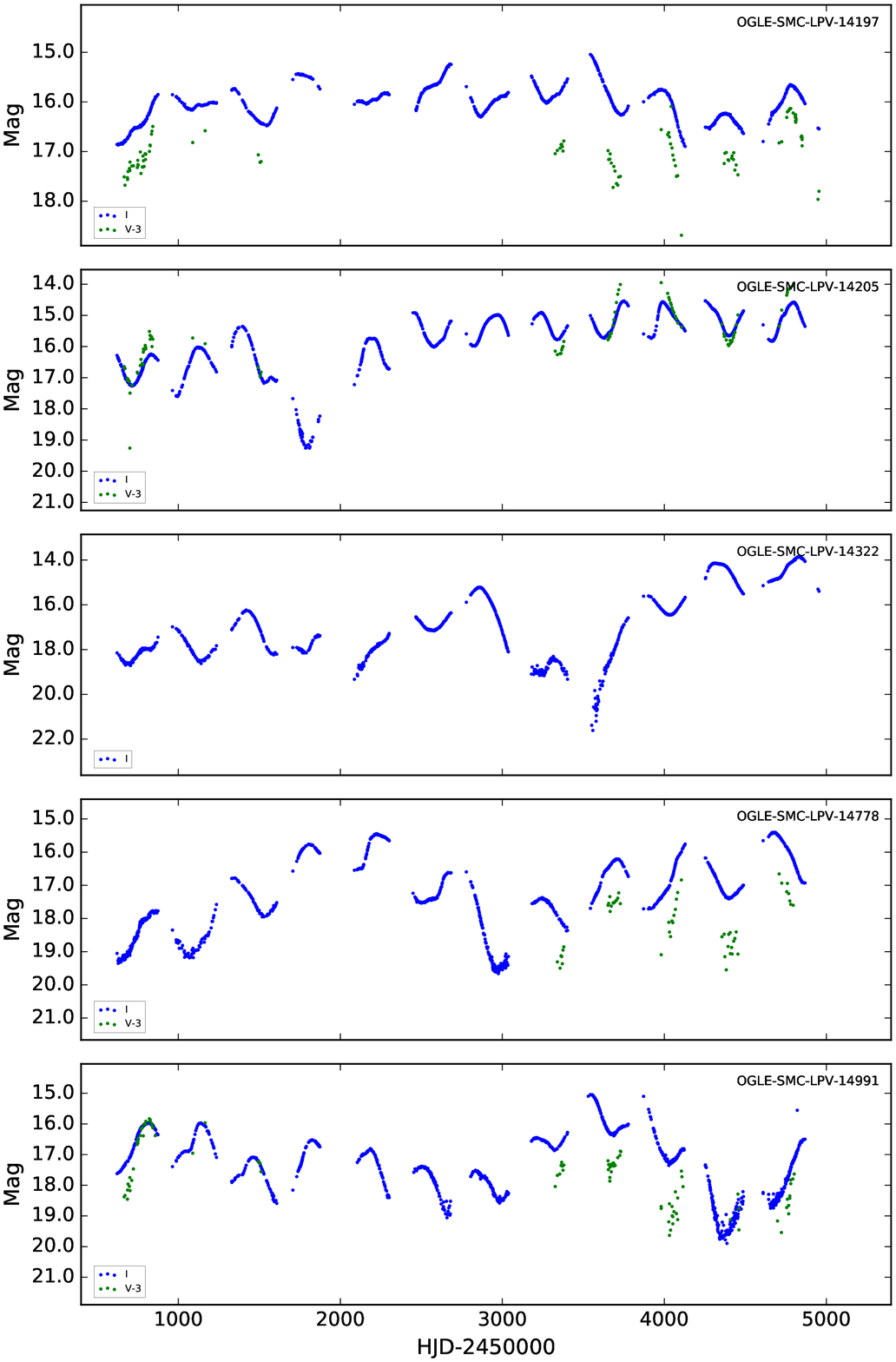}{Light curves in $I$ ({\em blue}) and $V$ ({\em green}) of new DY~Per candidates in the SMC.}
\label{fig:fig12}
\vskip 1cm

\IBVSfig{20cm}{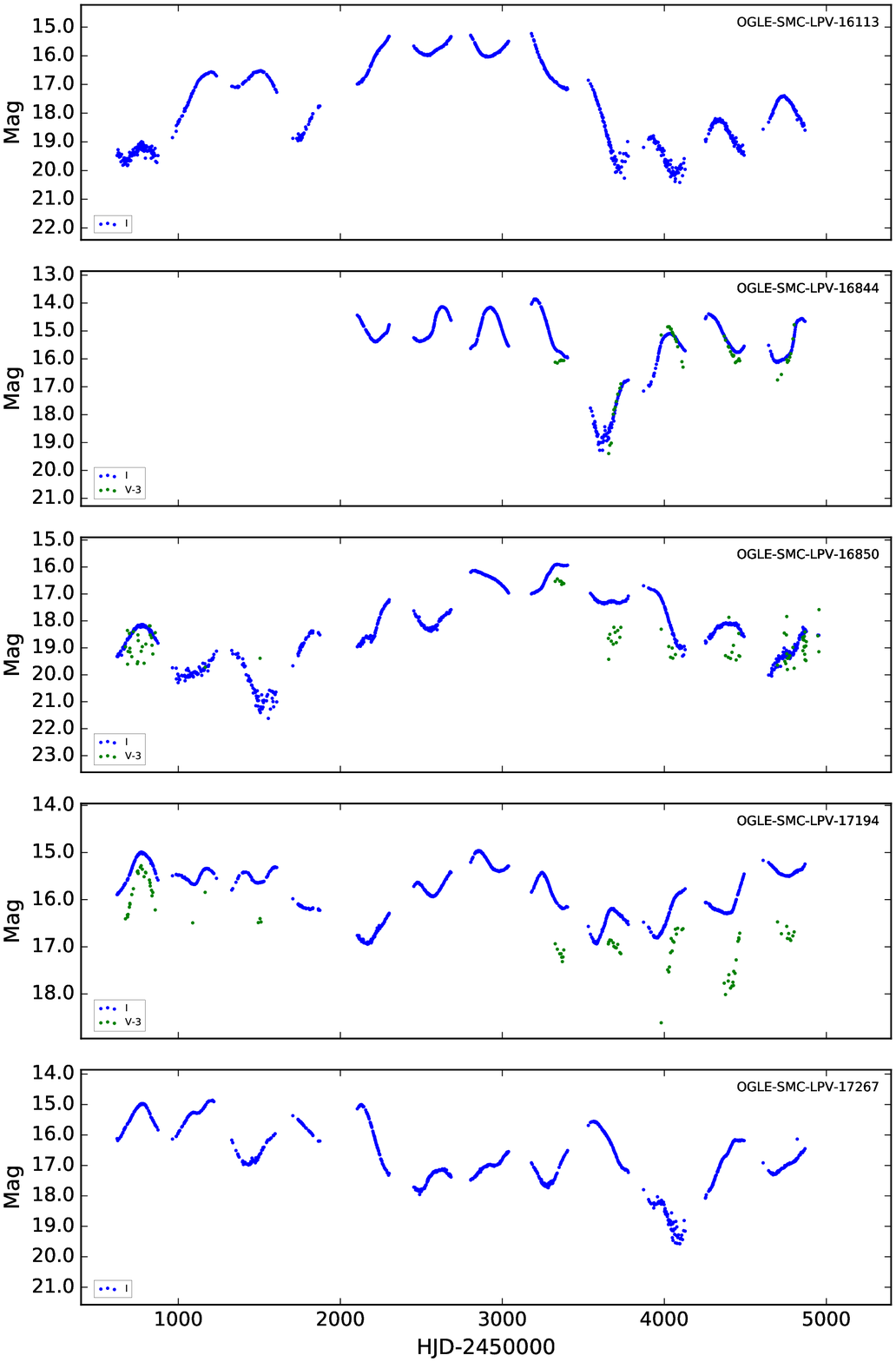}{Light curves in $I$ ({\em blue}) and $V$ ({\em green}) of new DY~Per candidates in the SMC.}
\label{fig:fig13}
\vskip 1cm

\IBVSfig{12cm}{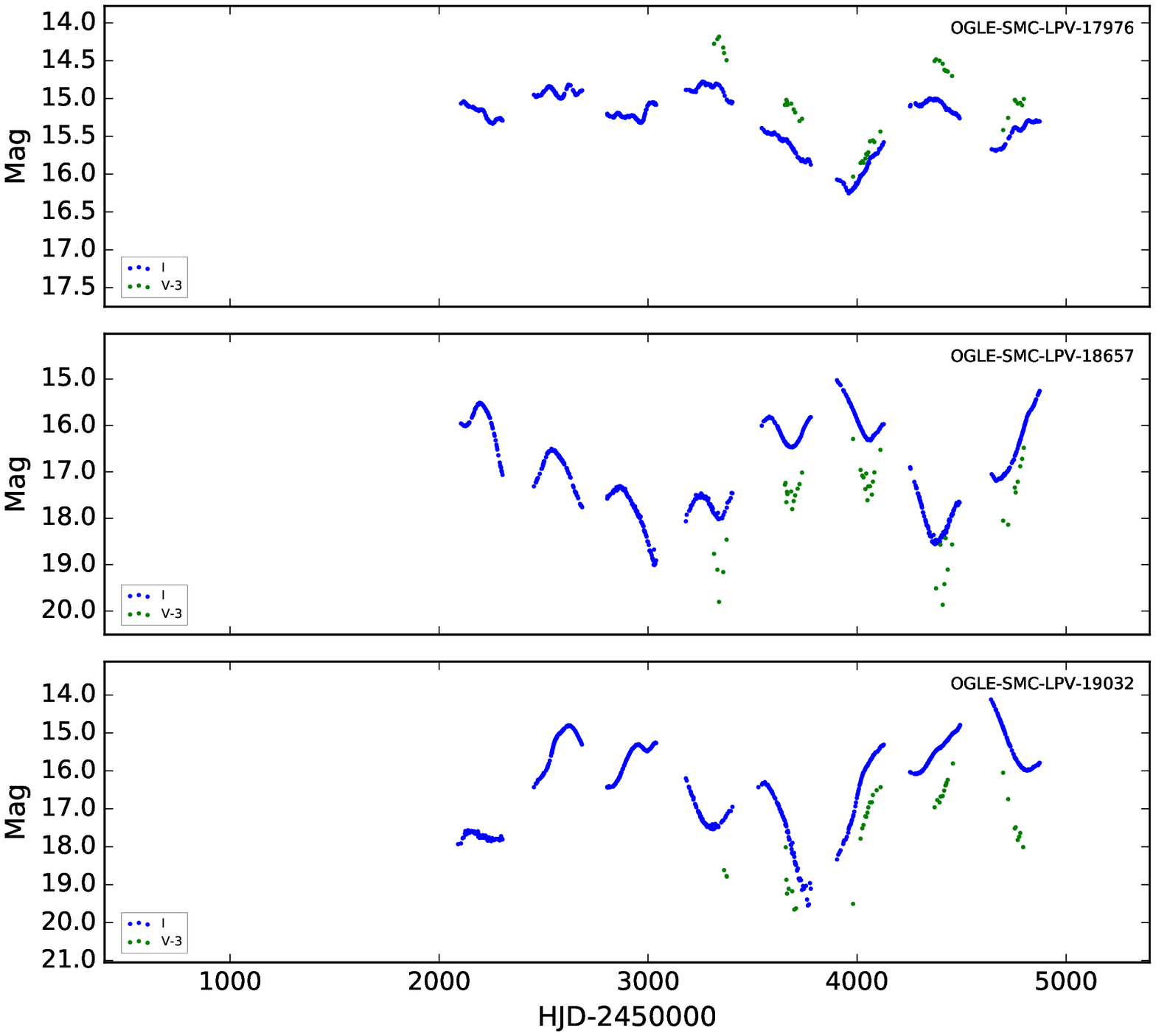}{Light curves in $I$ ({\em blue}) and $V$ ({\em green}) of new DY~Per candidates in the SMC.}
\label{fig:fig14}
\vskip 1cm
\IBVSfig{12cm}{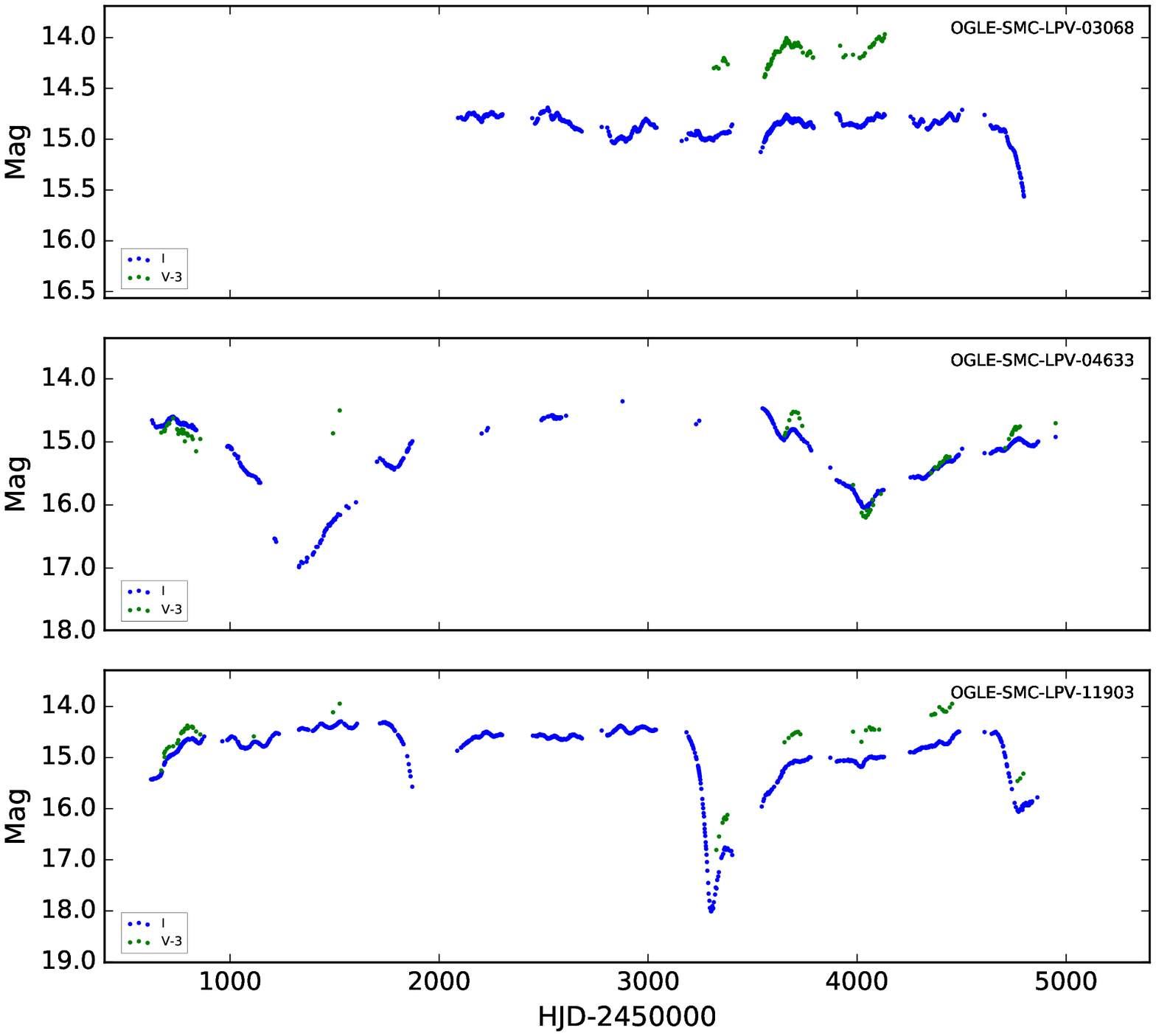}{Light curves in $I$ ({\em blue}) and $V$ ({\em green}) of previously confirmed DY~Per stars in the SMC (Tisserand et al. 2009), identified in this paper using OGLE data. Note the RCB-like light curve shape of OGLE-SMC-LPV-11903.}
\label{fig:fig15}
\vskip 1cm

\IBVSfig{15cm}{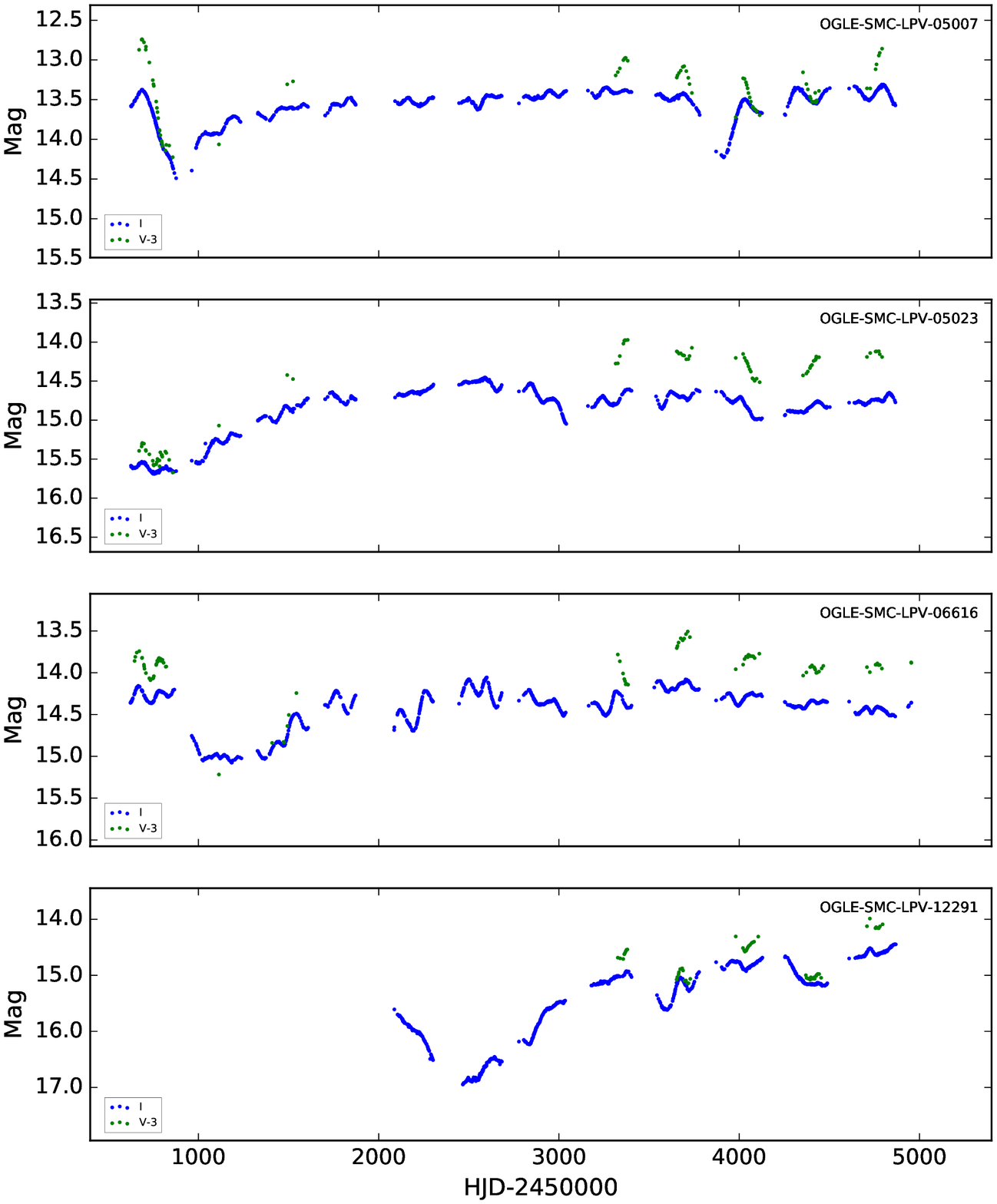}{Light curves in $I$ ({\em blue}) and $V$ ({\em green}) of candidate DY~Per stars in the SMC (Tisserand et al. 2009), identified in this paper using OGLE data. Note that we include in this plot the ``borderline'' DY~Per-like star OGLE-SMC-LPV-05007 (see text for details). 
}
\label{fig:fig16}
\vskip 1cm
\IBVSfig{8cm}{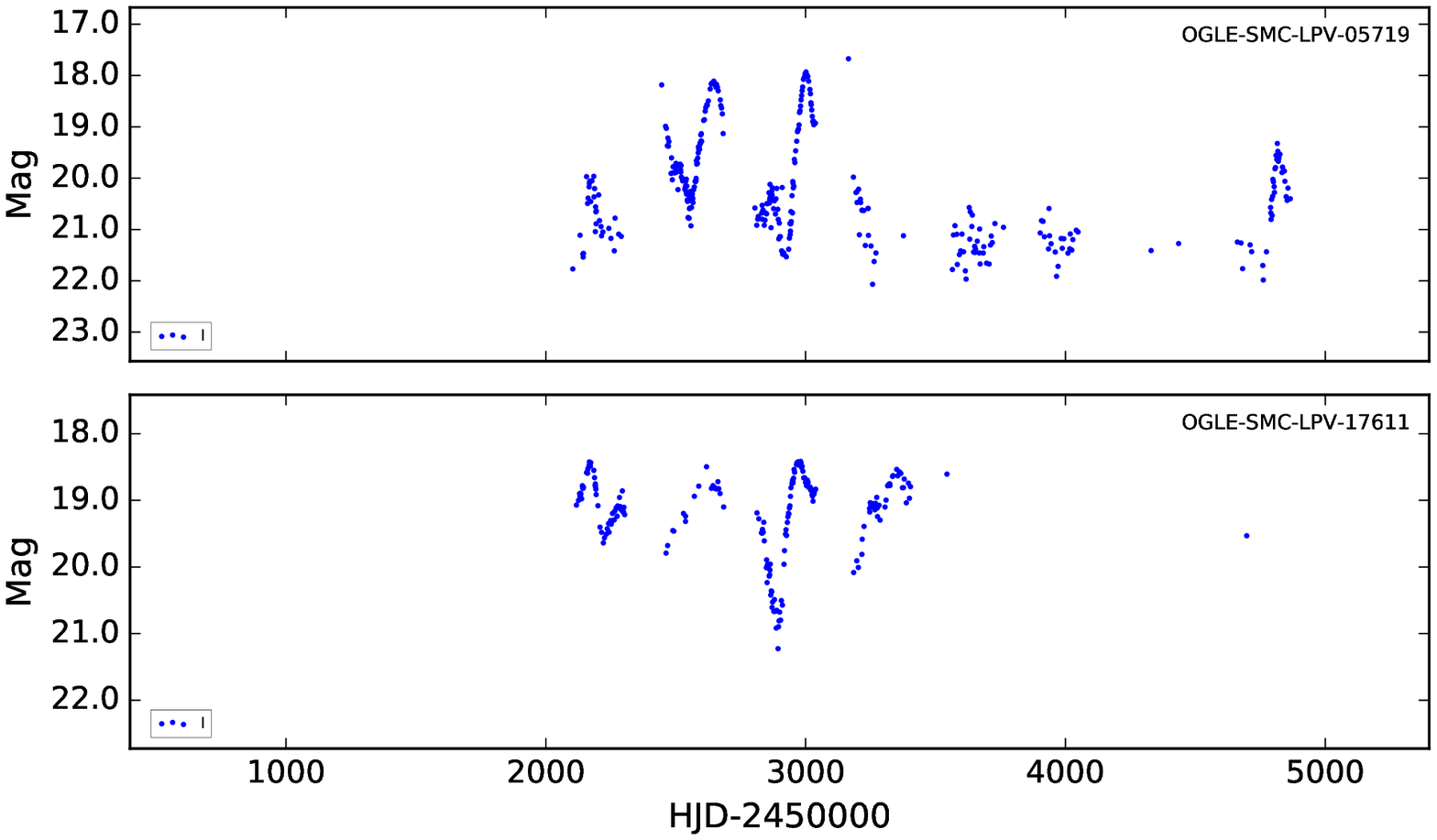}{({\em Upper panel}): light curve in $I$ ({\em blue}) of a known RCB candidate in the SMC (MSX-SMC-014) (Kraemer et al. 2005), identified in this paper as OGLE-SMC-LPV-05719. ({\em Bottom panel}): our new RCB candidate OGLE-SMC-LPV-17611.}
\label{fig:fig17}
\vskip 1cm 

\end{document}